\documentclass[aps,amsmath,amssymb,prd,showpacs,floatfix,preprint,superscriptaddress,nofootinbib,12pt]{article}
\pdfoutput=1
\usepackage{jheppub}
\usepackage{bm}
\usepackage{ulem}
\allowdisplaybreaks[3]

\begin{document}

\begin{flushright}
\small{SU-ITP-15/12}
\end{flushright}

\title{Dissecting holographic conductivities}

\author[1]{Richard A. Davison}
\affiliation[1]{Lorentz Institute for Theoretical Physics, Niels Bohrweg 2, Leiden NL-2333 CA, The Netherlands}
\author[2,3]{and Blaise Gout\'{e}raux}
\affiliation[2]{Stanford Institute for Theoretical Physics, Department of Physics, Stanford University,
Stanford, CA 94305, USA}
\affiliation[3]{APC, Universit\'e Paris 7, CNRS, CEA, Observatoire de Paris, Sorbonne Paris Cit\'e, F-75205, Paris Cedex 13, France}
\emailAdd{davison@lorentz.leidenuniv.nl, gouterau@stanford.edu}

\abstract{
The DC thermoelectric conductivities of holographic systems in which translational symmetry is broken can be efficiently computed in terms of the near-horizon data of the dual black hole. By calculating the frequency dependent conductivities to the first subleading order in the momentum relaxation rate, we give a physical explanation for these conductivities in the simplest such example, in the limit of slow momentum relaxation. Specifically, we decompose each conductivity into the sum of a coherent contribution due to momentum relaxation and an incoherent contribution, due to intrinsic current relaxation. This decomposition is different from those previously proposed, and is consistent with the known hydrodynamic properties in the translationally invariant limit. This is the first step towards constructing a consistent theory of charged hydrodynamics with slow momentum relaxation.
}

\maketitle

\section{Introduction}

Understanding the transport properties of strongly interacting many-body systems with no quasiparticles is a topic of much interest for both experimental and theoretical reasons. One class of theoretical examples are the strongly interacting quantum field theories which are holographically dual to classical theories of gravity. Holographic duality can be exploited to calculate the transport properties of these examples in a relatively simple way, with the goal of determining non-holographic effective theories which control these properties. Recent examples of this approach include \cite{Hartnoll:2007ih,Hartnoll:2012rj,Davison:2013txa,Lucas:2014zea,Blake:2014yla,Hartnoll:2014lpa,Davison:2014lua,Lucas:2015pxa,Hartnoll:2015sea}. In this paper, we study holographic systems with weakly broken translational invariance as a first step in formulating a general hydrodynamic theory of strongly interacting systems with slow momentum relaxation.

The transport properties of primary interest are the electrical ($\sigma$), thermoelectric ($\alpha$) and thermal ($\bar{\kappa}$) conductivities that control the linear response of the electric current $J$ and the heat current $Q$ to small electric fields $E$ and temperature gradients $\nabla T$
\begin{equation}
\label{eq:transportmatrixintroduction}
\left(\begin{array}{c}
J\\
Q\end{array} \right) = \left(
\begin{array}{cc}
\sigma & \alpha T \\
\alpha T& \bar\kappa T
\end{array} \right)\left(\begin{array}{c}
E\\
-\nabla T/T\end{array} \right). 
\end{equation}
The primary consideration in determining the qualitative form of these conductivities in holographic systems is whether the total momentum $ P$ of the system is approximately conserved or not. In this paper, we will primarily address situations in which this is the only long-lived quantity. In these cases, a perturbative expansion in the (small) momentum relaxation rate $\Gamma$ can be performed within the memory matrix formalism \cite{forster1990hydrodynamic,Hartnoll:2007ih,Hartnoll:2012rj,Mahajan:2013cja,Lucas:2015pxa}. To leading order in this expansion, the conductivities are all Drude-like, with DC values determined by $\Gamma$ and by the static susceptibilities $\chi_{JP}$ and $\chi_{QP}$ of the translationally invariant state where momentum is exactly conserved:
\begin{equation}
\label{eq:memmatrixacresults}
\sigma\left(\omega\right)=\frac{\chi_{JP}^2}{\chi_{PP}}\frac{1}{\Gamma-i\omega},\;\;\;\;\;\;\;\;\alpha\left(\omega\right)=\frac{\chi_{JP}\chi_{QP}}{T\chi_{PP}}\frac{1}{\Gamma-i\omega},\;\;\;\;\;\;\;\;\bar{\kappa}\left(\omega\right)=\frac{\chi_{QP}^2}{T\chi_{PP}}\frac{1}{\Gamma-i\omega}.
\end{equation}
Physically, any current $ A$ which overlaps with the momentum ($\chi_{AP}\ne0$) cannot decay at a rate larger than $\Gamma$ at late times. The slow relaxation of momentum acts as a bottleneck that forces the current into a \textit{coherent} late time response, even if its intrinsic relaxation timescale is fast. The low energy optical conductivity is dominated by a single pole that is parametrically close to the origin. In the opposite situation, when a current does not overlap with the momentum ($\chi_{AP}=0$), it will dissipate at its intrinsic rate. This is an example of \textit{incoherent} transport \cite{Hartnoll:2014lpa} and is the case for the electric conductivity in charge conjugation symmetric states, for example. 

These results form a basic, non-holographic effective theory that describes the transport of charge and energy in holographic systems in which momentum is approximately conserved. For these systems, one can attempt to enhance this basic effective theory by combining it with our knowledge of the system's properties in the translationally invariant limit, in which its late time behaviour is described by the laws of relativistic (conformal) hydrodynamics. The simplest way to incorporate the above results from the memory matrix formalism is to modify the momentum conservation equation in hydrodynamics, such that $P$ decays at a constant rate $\Gamma$. This yields the conductivities \cite{Hartnoll:2007ih}
\begin{equation}
\begin{aligned}
\label{eq:hydroacresult}
\sigma\left(\omega\right)&=\frac{\chi_{JP}^2}{\chi_{PP}}\frac{1}{\Gamma-i\omega}+\sigma_Q=\frac{n^2}{\epsilon+p}\frac{1}{\Gamma-i\omega}+\sigma_Q,\\
\alpha\left(\omega\right)&=\frac{\chi_{JP}\chi_{QP}}{T\chi_{PP}}\frac{1}{\Gamma-i\omega}-\frac{\mu}{T}\sigma_Q=\frac{n s}{\epsilon+p}\frac{1}{\Gamma-i\omega}-\frac{\mu}{T}\sigma_Q,\\
\bar{\kappa}\left(\omega\right)&=\frac{\chi_{QP}^2}{T\chi_{PP}}\frac{1}{\Gamma-i\omega}+\frac{\mu^2}{T}\sigma_Q=\frac{s^2T}{\epsilon+p}\frac{1}{\Gamma-i\omega}+\frac{\mu^2}{T}\sigma_Q,
\end{aligned}
\end{equation}
where $\epsilon$, $p$, $n$, $\mu$ and $s$ are the energy density, pressure, charge density, chemical potential, and entropy density of the state respectively. Each conductivity has a coherent contribution at leading order in $\Gamma$, as well as a subleading incoherent contribution proportional to the intrinsic conductivity of the hydrodynamic state $\sigma_Q$.\footnote{Note that $\sigma_Q$ is an intrinsic property of the \textit{finite} density state obtained by perturbing the neutral UV CFT by a chemical potential.}  
The former is in perfect agreement with the memory matrix results (\ref{eq:memmatrixacresults}), while the latter is a correction due to long-lived diffusive modes, whose form is specified precisely by the relativistic hydrodynamic theory in terms of a single transport coefficient $\sigma_Q$. The memory matrix results (\ref{eq:memmatrixacresults}) can be extended to incorporate the effects of diffusion in more general setups \cite{Lucas:2015pxa} -- these are independent of the momentum relaxation rate and enter at the first subleading order in a small $\Gamma$ expansion.

Recent advances in the holographic description of strongly interacting systems with momentum relaxation, in particular, the discovery of analytically tractable toy models of such systems \cite{Vegh:2013sk,Donos:2013eha,Andrade:2013gsa}, and the development of efficient calculational tools to determine their DC conductivities \cite{Blake:2013bqa,Donos:2014uba,Gouteraux:2014hca,Donos:2014cya,Donos:2014oha}, have made it easy to test this modified version of hydrodynamics. At leading order in $\Gamma$ \cite{Horowitz:2012ky,Davison:2013jba,Davison:2014lua,Lucas:2015vna}, the holographic results are consistent with those of the memory matrix (\ref{eq:memmatrixacresults}) and therefore with the leading order modified hydrodynamic results (\ref{eq:hydroacresult}). However, the holographic results are inconsistent with the modified hydrodynamic results at subleading order. To be explicit, we will consider the gravitational action \cite{Andrade:2013gsa}
\begin{equation}
\label{eq:axiontheoryaction}
S=\int d^4x \sqrt{-g}\left(\mathcal{R}+6-\frac{1}{4}F_{\mu\nu}F^{\mu\nu}-\frac{1}{2}\sum_{I=1}^{2}\partial^\mu\phi_I\partial_\mu\phi_I\right),
\end{equation}
which has the analytic black brane solution \cite{Bardoux:2012aw}
\begin{align}
\label{eq:axiontheorymetricsolution}
ds^2&=-r^2f(r)dt^2+r^2\left(dx^2+dy^2\right)+\frac{dr^2}{r^2f(r)},\;\;\;\;\;\;\phi_1=mx,\;\;\;\;\;\;\phi_2=my,\\
f(r)&=1-\frac{m^2}{2r^2}-\frac{r_0^3}{r^3}\left(1-\frac{m^2}{2r_0^2}+\frac{\mu^2}{4r_0^2}\right)+\frac{\mu^2r_0^2}{4r^4},\;\;\;\;\;\;\;A_t(r)=\mu\left(1-\frac{r_0}{r}\right). \nonumber
\end{align} 
The massless scalar fields $\phi_I$ break translational symmetry and so the DC conductivities are finite \cite{Andrade:2013gsa,Donos:2014cya}
\begin{equation}
\label{eq:hologDCresults}
\sigma_{DC}=1+\frac{\mu^2}{m^2},\;\;\;\;\;\;\;\;\;\;\alpha_{DC}=\frac{4\pi n}{m^2},\;\;\;\;\;\;\;\;\;\;\bar{\kappa}_{DC}=\frac{4\pi sT}{m^2}\,,
\end{equation}
where the explicit expressions for the energy, charge and entropy density of this state are given in \eqref{eq:thermo1} and \eqref{eq:thermo2}. 
The parameter $m$ controls the strength of translational symmetry breaking and therefore the rate of momentum relaxation in the dual field theory state. 

In \cite{Blake:2013bqa,Gouteraux:2014hca,Blake:2014yla,Donos:2014cya,Donos:2014oha}, it was suggested that $\sigma_{DC}$ could be interpreted as being composed of two physically distinct pieces: a coherent contribution $\mu^2/m^2$ due to momentum relaxation, and an incoherent contribution $1$ (see \cite{Kim:2014bza,Amoretti:2014zha} for further related work on the frequency dependence of the thermoelectric conductivities). However, this is inconsistent with the known value of the incoherent contribution $\sigma_Q$ in the translationally invariant ($m=0$) limit \cite{Hartnoll:2007ip} 
\begin{equation}
\label{eq:rnads4sigmaq}
\sigma_Q=\left(\frac{sT}{3\epsilon/2}\right)^2\Biggr|_{m=0}=\left.\left(\frac{3-\frac{\mu^2}{4r_0^2}}{3\left(1+\frac{\mu^2}{4r_0^2}\right)}\right)^2\right|_{m=0}.
\end{equation}
Furthermore, it is clear that a decomposition of this kind is inconsistent with the other hydrodynamic DC conductivities, as can be seen by comparing \eqref{eq:hologDCresults} with the DC limit of \eqref{eq:hydroacresult}.

In this paper we resolve these problems, and for the first time provide a clear description of the physical processes underlying the simple DC conductivities (\ref{eq:hologDCresults}), by analytically calculating the low frequency conductivities for the holographic theory (\ref{eq:axiontheorymetricsolution}) at small values of $m$ where there is approximate momentum conservation. We identify two physically distinct contributions to each conductivity -- a coherent contribution controlled by the slow relaxation of momentum, and an incoherent contribution due to the intrinsic conductivity $\sigma_Q$. The value of $\sigma_Q$ we obtain is consistent with the known value in the translationally invariant limit \eqref{eq:rnads4sigmaq} \cite{Hartnoll:2007ip}. Technically, we achieve this decomposition by changing basis from the currents $J$ and $Q$ to more theoretically convenient currents $J_\pm$ which are orthogonal: the two-point retarded Green's function of $J_+$ with $J_-$ vanishes. To the first subleading order at small $\Gamma$ the conductivity of $J_-$ is entirely coherent, and that of $J_+$ is entirely incoherent. In the strict $\Gamma=0$ limit, these reduce to the currents $P$ and $3\epsilon_0 J/2-n_0P$ respectively, which decouple and capture the entirely coherent and entirely incoherent responses respectively in the translationally invariant, hydrodynamic system (subscript 0s here denote the thermodynamic quantities of the $m=0$ state).

To the first subleading order at small $\omega$ and $\Gamma$, with $\omega/\Gamma$ fixed, we find that the frequency dependent conductivities take the form
\begin{equation}
\begin{aligned}
\label{eq:OurResults}
\sigma\left(\omega\right)&=\frac{\frac{\mu^2}{m^2}+(1-\sigma_Q)+O(\omega,\Gamma)}{1-i\omega/\Gamma}+\sigma_Q+O(\omega,\Gamma),\\
\alpha\left(\omega\right)&=\frac{\frac{4\pi n}{m^2}+\frac{\mu}{T}\sigma_Q+O(\omega,\Gamma)}{1-i\omega/\Gamma}-\frac{\mu}{T}\sigma_Q+O(\omega,\Gamma),\\
\bar{\kappa}\left(\omega\right)&=\frac{\frac{4\pi sT}{m^2}-\frac{\mu^2}{T}\sigma_Q+O(\omega,\Gamma)}{1-i\omega/\Gamma}+\frac{\mu^2}{T}\sigma_Q+O(\omega,\Gamma),
\end{aligned}
\end{equation}
where the momentum relaxation rate is
\begin{equation}
\Gamma=\frac{sm^2}{4\pi (\epsilon+p)}\left(1+\lambda m^2+O(m^4)\right),
\end{equation}
the thermodynamic quantities are those of the $m\ne0$ state, and $\lambda$ is given in equation \eqref{eq:GammaExp2}. For comparison with the hydrodynamic results (\ref{eq:hydroacresult}), these expressions may be written as
\begin{align}
\label{eq:OurResultsHydroForm}
\sigma\left(\omega\right)&=\frac{\frac{n^2}{\epsilon+p}+\Gamma\left(1-\sigma_Q+\lambda\mu^2\right)+O(\Gamma^2,
\omega\Gamma,\omega^2)}{\Gamma-i\omega}+\sigma_Q+O(\omega,\Gamma),\\
\alpha\left(\omega\right)&=\frac{\frac{n s}{\epsilon+p}+\Gamma\left(\frac{\mu}{T}\sigma_Q+4\pi n\lambda\right)+O(\Gamma^2,
\omega\Gamma,\omega^2)}{\Gamma-i\omega}-\frac{\mu}{T}\sigma_Q+O(\omega,\Gamma),\\
\bar{\kappa}\left(\omega\right)&=\frac{\frac{s^2T}{\epsilon+p}+\Gamma\left(-\frac{\mu^2}{T}\sigma_Q+4\pi s T\lambda\right)+O(\Gamma^2,
\omega\Gamma,\omega^2)}{\Gamma-i\omega}+\frac{\mu^2}{T}\sigma_Q+O(\omega,\Gamma).
\end{align}

These results isolate the reason for the inconsistency between the modified version of hydrodynamics and the holographic system: the modified version of hydrodynamics does not adequately describe the coherent component of the system's response. Although it reproduces the correct coherent contribution at leading order in $\Gamma$, it does not adequately account for the first subleading corrections to this. These corrections are important as they enter at the same order as the incoherent contribution, and emphasize the need for a more systematic derivation of how hydrodynamics is modified by the weak breaking of translational symmetry. Our calculation also highlights the important message that it is in general not possible to separate the coherent and incoherent contributions to the conductivities from their DC expressions (\ref{eq:hologDCresults}) alone.
We note that the obvious decomposition of the DC conductivities \eqref{eq:hologDCresults} still has physical meaning in terms of the DC conductivities at zero electric or heat current \cite{Mahajan:2013cja,Donos:2014cya}.

Finally, although the main focus of our paper is the limit of slow momentum relaxation, we can also easily access the regime of fast momentum relaxation in the holographic theory (\ref{eq:axiontheoryaction}). This is a regime in which neither the hydrodynamic (\ref{eq:hydroacresult}) nor memory matrix results (\ref{eq:memmatrixacresults}) are applicable. For \textit{any} value of $m$, it is possible to diagonalise the response of the currents by changing to an appropriate basis $J_\pm$. In the limit $m\rightarrow\infty$, the decoupled currents are precisely $J$ and $Q$, the electrical and heat currents. It would be very interesting to determine whether a low energy decoupling of this type is present more generally in systems with fast momentum relaxation (in particular, those with a potential for the scalar fields, which are more reliable from the point of view of string theory).

In Section \ref{section:diag}, we identify the diagonal $J_\pm$ basis of currents in the field theory by decoupling the bulk field equations, and examine how these decoupled currents relate to $J$ and $Q$ in various limits of interest. In Section \ref{section:freq}, we determine the frequency dependence of the conductivities of the currents $J_\pm$ in the limit of slow momentum relaxation, showing that one is coherent and that one is incoherent, and explain what this means for the conductivities of $J$ and $Q$. We conclude in Section \ref{section:outlook} with an outlook for future work. The appendices contain some technical details of our holographic Green's function calculations.

\section{Diagonalisation of the conductivities\label{section:diag}}

To determine the frequency dependent thermoelectric conductivities in the strongly interacting field theory state dual to (\ref{eq:axiontheorymetricsolution}), we will use the Kubo formul\ae\ \cite{Hartnoll:2007ih} which relate these conductivities to the retarded two-point functions $G^R$ of the currents $J$ and $Q\equiv J_E-\mu J$, where $J_E$ is the energy current:
\begin{equation}
\begin{aligned}
\sigma\left(\omega\right)&=\frac{i}{\omega}\left[G^R_{JJ}\left(\omega,k=0\right)-G^R_{JJ}\left(\omega=0,k\rightarrow0\right)\right],\\
\alpha\left(\omega\right)&=\frac{i}{\omega T}\left[G^R_{QJ}\left(\omega,k=0\right)-G^R_{QJ}\left(\omega=0,k\rightarrow0\right)\right],\\
\bar{\kappa}\left(\omega\right)&=\frac{i}{\omega T}\left[G^R_{QQ}\left(\omega,k=0\right)-G^R_{QQ}\left(\omega=0,k\rightarrow0\right)\right].
\end{aligned}
\label{eq:thermoelecgreensfns}
\end{equation}
To evaluate the Green's functions on the right hand side, it is convenient to first change the basis of currents and not work directly with $J$ and $Q$, as we will shortly describe. Following that, we will use the standard tools of the AdS/CFT correspondence to compute the Green's functions.

\subsection{Decoupling of the gravitational equations of motion}

To determine the two-point functions of $J$ and $Q$, we consider the following consistent set of linear perturbations around the black brane solution (\ref{eq:axiontheorymetricsolution}) 
\begin{equation}
\label{eq:quadraticperturbationsdefinition}
\begin{aligned}
\delta g_{t}^y(t,r)&=\int\frac{d\omega}{2\pi}h^y_t(r,\omega)e^{-i\omega t}\,,\quad\delta g_{r}^y(t,r)=\int\frac{d\omega}{2\pi}h^y_r(r,\omega)e^{-i\omega t}\,,\\
\delta A_y(t,r)&=\int\frac{d\omega}{2\pi}a_y(r,\omega) e^{-i\omega t}\,,\quad \delta\phi_2(r,t)=\int\frac{d\omega}{2\pi}\chi_2(r,\omega) e^{-i\omega t},
\end{aligned}
\end{equation}
where indices are raised with the background metric. These obey the following linearised equations of motion (where primes denote derivatives with respect to $r$)
\begin{equation}
\begin{split}
0=&\left(r^2f a_y'\right)'+\frac{\omega^2}{r^2f}a_y+r^2 A_t' {h_t^y}'+i\omega r^2 A_t'h_r^y,\\
0=&\frac1{r^2}\left(r^4 {h_t^y}'\right)'+\frac{i\omega}{r^2}\left(r^4 h_r^y\right)'+h_t^y\left(6-\frac6{f}+\frac{A_t'^2}{2f}+\frac{2r f'}{f}\right)+A_t' a_y'-\frac{i m \omega  }{r^2 f}\chi_2,\\
0=&\frac{i \omega   A_t'}{2 r^2 f}a_y-\frac{1}{4} r^2 \left(A_t'^2+4 r f'+12 f-12+\frac{2\omega ^2 }{r^2 f}\right)h_r^y+\frac{i \omega  }{2 f}{h_t^y}'-\frac{m}2\chi_2',\\
0=&\frac1{r^2}\left(r^4f\chi_2'\right)'-\frac{m}{r^2}\left(r^4 f h_r^y\right)'+\frac{\omega^2}{r^2f}\chi_2-\frac{i m\omega}{r^2f}h_t^y,
\end{split}
\end{equation}
which are comprised of two linearly independent dynamical equations, and one constraint equation.

We can decouple the two dynamical equations by changing variables to
\begin{equation}
\label{eq:decoupledfieldeqns}
\frac{d}{dr}\left[r^2f\psi_\pm'\right]+\frac{1}{r^2f}\left(\omega^2-m^2f+r^3ff'+m\gamma_\pm A_t'rf\right)\psi_\pm=0,
\end{equation}
where
\begin{equation}
\label{eq:defnofdecoupledbulkfields}
\psi_\pm\equiv\frac{1}{m}\left[r^3\left({h^y_t}'+i\omega h^y_r\right)+rA_t'a_y\right]+\gamma_\pm a_y\,.
\end{equation}
Here
\begin{equation}
\gamma_\pm\equiv-\frac{3\epsilon}{4mn}\left(1\pm\sqrt{1+\frac{16m^2n^2}{9\epsilon^2}}\right),
\end{equation}
with $\epsilon$ and $n$, the energy and charge densities of the state, given by
\begin{equation}
\label{eq:thermo1}
\epsilon=2r_0^3\left(1-\frac{m^2}{2r_0^2}+\frac{\mu^2}{4r_0^2}\right)\,,\quad\quad\quad n=\mu r_0.
\end{equation}
The other thermodynamic properties of the equilibrium state are \cite{Andrade:2013gsa}
\begin{equation}
\label{eq:thermo2}
4\pi T=3 r_0-\frac{m^2}{2r_0}-\frac{\mu^2}{4 r_0}\,, \qquad s=4\pi r_0^2\,,\qquad p=\langle T^{ii}\rangle+r_0 m^2=\frac{1}{2}\epsilon+r_0m^2.
\end{equation}

This change of variables corresponds, in the field theory, to a change of operator basis\footnote{The energy current in our system is $J_E\equiv T^{tx}(k=0)$, which is the momentum when $m=0$.} from $(J_E,J)$ to $(J_+,J_-)$. The decoupling of these variables in the bulk corresponds to a diagonalisation of the matrix of two-point functions of the dual operators i.e.~it corresponds to diagonalising the matrix of conductivities. By a careful analysis of the on-shell action (see appendix \ref{app:onshellactionappendix}), we find that the two-point retarded Green's functions $G^R$ are related to the boundary behaviour of the decoupled fields $\psi_\pm$ via
\begin{equation}
\label{eq:subtractedbasiccorrs}
\begin{aligned}
&\langle JJ\rangle(\omega)=-\frac{1}{\gamma_+-\gamma_-}\left(\gamma_+\Theta_+\left(\omega\right)-\gamma_-\Theta_-\left(\omega\right)\right),\\
&\langle JJ_E\rangle(\omega)=\langle J_EJ\rangle(\omega)=-\frac{m}{\gamma_+-\gamma_-}\left(\Theta_+\left(\omega\right)-\Theta_-\left(\omega\right)\right)+n,\\
&\langle J_EJ_E \rangle(\omega)=-\frac{m^2}{\gamma_+-\gamma_-}\left(\gamma_+\Theta_-\left(\omega\right)-\gamma_-\Theta_+\left(\omega\right)\right)+\frac{3}{2}\epsilon,
\end{aligned}
\end{equation}
where the angled brackets denote the specific combinations of retarded two-point Green's functions which enter in the Kubo formul\ae\ for conductivities \eqref{eq:thermoelecgreensfns}
\begin{equation}
\label{eq:subtractedcorrelatordefinition}
\langle\mathcal{O}\mathcal{O}\rangle(\omega)\equiv G^R_{\mathcal{O}\mathcal{O}}(\omega,k=0)-G^R_{\mathcal{O}\mathcal{O}}(\omega=0,k\rightarrow0),
\end{equation}
and where
\begin{equation}
\Theta_\pm\left(\omega\right)=-r^2\frac{\psi_\pm'}{\psi_\pm}\Biggr|_{r\rightarrow\infty},
\end{equation}
are determined by solving the decoupled equations of motion with ingoing boundary conditions at the horizon, and contain all of the frequency dependence of the correlators.

It is simple to diagonalise the matrix of correlators by defining the currents (with overall normalisation constants $a_\pm$)
\begin{equation}
\label{eq:defofdecoupledcurrents}
J_\pm=a_\pm\left(J_E+\gamma_\pm mJ\right),
\end{equation}
so that $\langle J_\pm J_\pm\rangle$ depends only on $\Theta_\pm$, and the cross correlator $\langle J_\pm J_\mp\rangle=0$. Physically, this diagonalisation of the matrix of conductivities means we can divide the response of the currents of our system, at any energy scale, into two completely independent sectors, each with its own spectrum of excitations. This situation is familiar, for example, in zero density, translationally invariant systems, where $J$ and $J_E$ decouple due to charge conjugation symmetry. In our case, there does not appear to be any symmetry protecting this exact decoupling at all energy scales, and we do not expect it to be true in general for holographic systems. The more pertinent question is whether the conductivity matrix can be diagonalised \textit{at low energies $\omega$} in more general holographic states. This does not necessarily require an exact decoupling of the bulk perturbations, and would be an indicator of the existence of a simple, low energy effective description of transport in these states. An example of this, when $m=0$, is described below.

The perturbation equations for this holographic model can also be completely decoupled at non-zero wavevectors $k$. Again, we expect that this feature is specific to this very simple example, and will not be true in general.

Inverting these relationships, we can express the responses of the correlators we are truly interested in -- those of the electrical and heat currents -- as linear combinations of those of the decoupled currents $J_\pm$ as follows:
\begin{equation}
\label{eq:condintermsdiagonalcurrents}
\begin{aligned}
&\langle JJ\rangle=\frac{1}{m^2}\left(\langle\mathcal{J}_+\mathcal{J}_+\rangle+\langle\mathcal{J}_-\mathcal{J}_-\rangle\right),\\
&\langle QJ\rangle=\langle JQ\rangle=-\frac{1}{m}\left[\left(\gamma_-+\frac{\mu}{m}\right)\langle\mathcal{J}_+\mathcal{J}_+\rangle+\left(\gamma_++\frac{\mu}{m}\right)\langle\mathcal{J}_-\mathcal{J}_-\rangle\right],\\
&\langle QQ\rangle=\left(\gamma_-+\frac{\mu}{m}\right)^2\langle\mathcal{J}_+\mathcal{J}_+\rangle+\left(\gamma_++\frac{\mu}{m}\right)^2\langle\mathcal{J}_-\mathcal{J}_-\rangle,\\
\end{aligned}
\end{equation}
where we have introduced the rescaled correlators
\begin{equation}
\label{eq:rescaledcorrelatorextractions}
\langle\mathcal{J}_\pm\mathcal{J}_\pm\rangle\left(\omega\right)\equiv\frac{\langle J_\pm J_\pm\rangle\left(\omega\right)}{a_\pm^2\left(\gamma_+-\gamma_-\right)^2}=\frac{\mp m^2\gamma_\pm}{\gamma_+-\gamma_-}\Theta_\pm\left(\omega\right)+\frac{1}{\left(\gamma_+-\gamma_-\right)^2}\left(\frac{3\epsilon}{2}+2m\gamma_\pm n\right),
\end{equation}
for convenience. From these, one simply needs to divide by the appropriate factor in the Kubo formul\ae\ (\ref{eq:thermoelecgreensfns}) to extract the relevant conductivity. The decoupled currents $\mathcal{J}_{\pm}$ are sourced by $\pm\left(E/m+(\gamma_\mp+\mu/m)\nabla T/T\right)$, and transport the conserved charge densities $\left(T^{tt}+\gamma_\pm m J^t\right)/\left(\gamma_+-\gamma_-\right)$.

\subsection{The decoupled currents in various limits}

Remarkably, we have managed to decouple the response of the currents $J_{\pm}$ at all frequencies, and for all values of the parameters $m,T$ and $\mu$. For certain values of the parameters, the decoupled currents $J_\pm$ take particularly simple forms, which allows us to ascribe a clear physical meaning to the decoupling. 

The simplest limit is already very familiar: when $\mu\rightarrow0$ at fixed $T$ and $m$, after an appropriate choice of normalisations the decoupled currents are $J_+\rightarrow J$ and $J_-\rightarrow J_E=Q$. This is simply the well-known result that at zero chemical potential, the heat and charge currents of a system decouple due to charge conjugation symmetry. The heat and charge conductivities may be qualitatively different from each other in this limit. The charge response will be incoherent as $J$ does not overlap with any almost conserved operators. As $Q$ overlaps with $P$, the heat response will be coherent when $P$ dissipates slowly (at small $m$), and incoherent otherwise.

There is another limit which is in fact rather similar to this: when $m\rightarrow\infty$ at fixed $T$ and $\mu$, after an appropriate choice of normalisations, the decoupled currents are just $J_+\rightarrow J$ and $J_-\rightarrow J_E-\mu J=Q$. This is a rather surprising result: in the limit of very strong translational symmetry breaking, the charge and heat currents decouple! Heuristically, it is as if there is an emergent form of charge conjugation symmetry in this limit. One way of understanding this is that when $m\gg\mu,T$, the contributions of the uncharged  scalar operators dual to the fields $\phi_I$ dominate the thermodynamic properties of the system such that it looks like a neutral state. In particular, the dimensionless ratio of charge density to entropy density, a thermodynamic measure of the ratio of charged to neutral degrees of freedom, approaches zero in this limit: $n/s\sim\mu/m\rightarrow0$. However, this is qualitatively different from the $\mu=0$ limit in that it is specifically the heat current $Q$ which decouples from $J$, while other neutral currents like $J_E$ still couple to $J$. It is clearly worth investigating to see if this a common feature of low energy transport in states of this type, or just a peculiarity of this holographic system. Finally, note that in contrast to the previous $\mu\rightarrow0$ limit, in this limit both the charge and heat conductivities will be incoherent, as momentum dissipates quickly in the limit $m\rightarrow\infty$.

Finally, there is the limit of slow momentum relaxation, in which we are mainly interested in the remainder of this paper. In the limit $m\rightarrow0$ with $T$ and $\mu$ fixed, the decoupled currents asymptote to
\begin{equation}
\label{eq:smallmdecoupledcurrents}
J_+\rightarrow J_E-\frac{3\epsilon_0}{2n_0}J+O(m^2),\;\;\;\;\;\;\;\;\;\; J_-\rightarrow J_E+O(m^2),
\end{equation}
after appropriate choices of normalisation, where the subscript $0$s denote the thermodynamic quantity in the $m=0$ state. This decoupling also has a clear physical origin, which is independent of holographic duality. In the strict $m=0$ limit, our state obeys the laws of conformal, relativistic hydrodynamics at low energies. In such a hydrodynamic state, the currents given in (\ref{eq:smallmdecoupledcurrents}) decouple at low energies. $J_E=P$ controls the coherent component of the system's response while the other current controls the incoherent response, since it decouples from the conserved total momentum $P$. See \cite{DGH} for more details.

\subsection{DC contributions of each sector \label{sec:DCconductivitiessection}}

The DC conductivities correspond to the $\omega\rightarrow0$ limits of the subtracted correlators, and are given analytically in (\ref{eq:hologDCresults}). We will confirm these results in the next section. From these, we can extract the DC limits of the diagonal correlators
\begin{equation}
\label{eq:DCdecoupledcorrelators}
\text{Im}\left[\lim_{\omega\rightarrow0}\frac{1}{\omega}\langle\mathcal{J}_\pm\mathcal{J}_\pm\rangle\left(\omega\right)\right]=\frac{1}{2}\left(m^2+\mu^2\right)\mp\frac{2n}{3\epsilon\sqrt{1+\frac{16m^2n^2}{9\epsilon^2}}}\left[\left(m^2+\mu^2\right)\left(\mu-\frac{3\epsilon}{4n}\right)+4\pi n T\right].
\end{equation}
Using the decomposition (\ref{eq:condintermsdiagonalcurrents}), we can then easily isolate how much each of the decoupled sectors contributes to each DC thermoelectric conductivity. Although the full DC conductivities are very simple, each individual contribution is given by a very complicated expression (which can be found by combining \eqref{eq:condintermsdiagonalcurrents} with \eqref{eq:DCdecoupledcorrelators}). The separation of the full conductivities into two decoupled sectors is highly non-trivial and cannot be guessed just from the form of the DC conductivities.

It is instructive to examine these contributions in the various limits of the previous subsection. We use the notation that the superscript $\pm$ indicates the contribution of the $J_\pm$ sector to each conductivity. In the $\mu\rightarrow0$ limit (at fixed $T,m$),
\begin{equation}
\begin{aligned}
\sigma_{DC}^+&\rightarrow\sigma_{DC}+O(\mu^{2}),\;\;\;\;\;\;\;\;\;\; \sigma_{DC}^-\rightarrow O(\mu^{2}),\;\;\;\;\;\;\;\;\;\; \\
\alpha_{DC}^+&\rightarrow O(\mu),\;\;\;\;\;\;\;\;\;\;\;\;\;\;\;\;\;\;\;\;\;\; \alpha_{DC}^-\rightarrow O(\mu),\;\;\;\;\;\;\;\;\;\; \\
\bar{\kappa}_{DC}^+&\rightarrow O(\mu^2),\;\;\;\;\;\;\;\;\;\;\;\;\;\;\;\;\;\;\;\;\; \bar{\kappa}_{DC}^-\rightarrow\bar{\kappa}_{DC}+O(\mu^2).\;\;\;\;\;\;\;\;\;\;
\end{aligned}
\end{equation}
This limit is well-known and the the nature of the decomposition is clear: this is the charge conjugation symmetric limit in which the decoupled currents $J_+$ and $J_-$ are the charge and heat currents respectively. Therefore, the electrical and thermal conductivities are controlled completely by the $+$ and $-$ sectors respectively, and the off-diagonal conductivity vanishes, to leading order at small $\mu$.

From this point of view, a qualitatively similar limit is the limit $m\rightarrow\infty$ (at fixed $T,\mu$), where
\begin{equation}
\begin{aligned}
\sigma_{DC}^+&\rightarrow\sigma_{DC}+O(m^{-4}),\;\;\;\;\;\;\;\;\;\; \sigma_{DC}^-\rightarrow O\left(m^{-4}\right),\;\;\;\;\;\;\;\;\;\; \\
\alpha_{DC}^+&\rightarrow O(m^{-1}),\;\;\;\;\;\;\;\;\;\;\;\;\;\;\;\;\;\;\;\; \alpha_{DC}^-\rightarrow O(m^{-2}),\;\;\;\;\;\;\;\;\;\; \\
\bar{\kappa}_{DC}^+&\rightarrow O(m^{-2}),\;\;\;\;\;\;\;\;\;\;\;\;\;\;\;\;\;\;\;\; \bar{\kappa}_{DC}^-\rightarrow\bar{\kappa}_{DC}+O(m^{-1}).\;\;\;\;\;\;\;\;\;\;
\end{aligned}
\end{equation}
Again, this is easy to understand: the decoupled currents in this limit are again the charge ($J_+$) and heat ($J_-$) currents, and so the electrical and thermal conductivities are finite\footnote{It is crucial here that the horizon radius is replaced by its expression in terms of physical parameters $T$, $\mu$ and $m$ before taking the $m\rightarrow\infty$ limit. When this is done, $\bar\kappa$ does not vanish at large $m$, contrarily to what the formula in \eqref{eq:hologDCresults} might appear to indicate.} and determined, at leading order, by the $+$ and $-$ sectors respectively, while the off-diagonal conductivities vanish at leading order in this limit.

Finally, let us turn to the limit we will address in the remainder of the paper: $m\rightarrow0$ (at fixed $T,\mu$). This is the limit of slow momentum relaxation. In this limit, the contribution of each of the sectors to the DC conductivities take a very suggestive form
\begin{equation}
\begin{aligned}
\sigma_{DC}^+&\rightarrow\sigma_Q+O(m^2),\;\;\;\;\;\;\;\;\;\;\;\;\;\;\;\;\;\;\; \sigma_{DC}^-\rightarrow\sigma_{DC}-\sigma_Q+O(m^2),\;\;\;\;\;\;\;\;\;\; \\
\alpha_{DC}^+&\rightarrow-\frac{\mu}{T}\sigma_Q+O(m^2),\;\;\;\;\;\;\;\;\;\;\;\;\; \alpha_{DC}^-\rightarrow\alpha_{DC}+\frac{\mu}{T}\sigma_Q+O(m^2),\;\;\;\;\;\;\;\;\;\; \\
\bar{\kappa}_{DC}^+&\rightarrow\frac{\mu^2}{T}\sigma_Q+O(m^2),\;\;\;\;\;\;\;\;\;\;\;\;\;\;\;\bar{\kappa}_{DC}^-\rightarrow\bar{\kappa}_{DC}-\frac{\mu^2}{T}\sigma_Q+O(m^2),
\end{aligned}
\label{eq:DCcondSmallM}
\end{equation}
where $\sigma_Q$ is given in equation (\ref{eq:rnads4sigmaq}). Each DC conductivity clearly decomposes into two independent pieces, one of which is exactly equal (to this order in $m^2$) to the incoherent contribution, due to intrinsic current relaxation, present in the effective hydrodynamic theory (\ref{eq:hydroacresult}). Although the full holographic DC conductivities are not consistent with this effective hydrodynamic theory, the decomposition above suggests that at least this part of the hydrodynamic theory is accurate. To confirm this, and give a more physical interpretation to the decoupling in this limit, we will now calculate the frequency dependence of each of the two independent contributions to the conductivities.

\section{Frequency dependent conductivities at small $m$ \label{section:freq}}

To determine the frequency dependent conductivities, we need to solve the perturbation equations (\ref{eq:decoupledfieldeqns}) at non-zero $\omega$, with ingoing boundary conditions at the black brane horizon. We can only find analytic solutions to these equations by working perturbatively in a small frequency expansion. The resulting conductivities we find, extracted via (\ref{eq:condintermsdiagonalcurrents}), are fractions, with both the numerator and denominator given by power series in $\omega$. This procedure was used in \cite{Policastro:2002se,Policastro:2002tn, Herzog:2002fn} to determine the two-point Green's functions of translationally invariant systems at small frequencies.

We begin by making the ansatze
\begin{equation}
\begin{aligned}
\label{eq:decoupledansatzeholo}
\psi_\pm(u)&=f(u)^{\frac{-i\tilde{\omega} r_0}{4\pi T}}\left(1-\frac{\mu\gamma_{\mp}}{m}u\right)\mathcal{F}_\pm(u),\\
\mathcal{F}_\pm(u)&=F_{\pm}^{(0)}(u)+\tilde{\omega} F_{\pm}^{(1)}(u)+\tilde{\omega}^2F_{\pm}^{(2)}(u)+O(\tilde{\omega}^3),
\end{aligned}
\end{equation}
for the gauge invariant fields, where we are using the dimensionless variables
\begin{equation}
u=\frac{r_0}{r}\,,\qquad \tilde{\omega}=\frac{\omega}{r_0}\,,\qquad \tilde m=\frac{m}{r_0}\,,\qquad \tilde\mu=\frac{\mu}{r_0}.
\end{equation}
We have factored out an oscillating function that corresponds to imposing ingoing boundary conditions at the black brane horizon, as well as an overall $u$-dependent function such that the leading terms $F_{\pm}^{(0)}(u)$ will be independent of $u$. To determine the functions $F_{\pm}^{(i)}(u)$, we substitute the ansatze (\ref{eq:decoupledansatzeholo}) into the equations of motion (\ref{eq:decoupledfieldeqns}) and expand as a power series in $\tilde{\omega}$. We then solve order-by-order for the functions $F_{\pm}^{(i)}(u)$, demanding that $\mathcal{F}_\pm(u)$ is regular and equal to a frequency-independent constant at the black brane horizon $u=1$. At leading order in $\tilde{\omega}$, we find that $F_{\pm}^{(0)}(u)=C_\pm$ is a constant which we will set to $1$ for convenience. At higher orders, $F_{\pm}^{(i)}$ are non-trivial functions of $u$ that satisfy equations of the form
\begin{equation}
{F_{\pm}^{(i)}}''(u)+g_{\pm}^{(i)}(u){F_{\pm}^{(i)}}'(u)=S_{\pm}^{(i)}(u).
\label{eq:diffeqschematicF}
\end{equation}
That is, they are first order, linear inhomogeneous equations for ${F_{\pm}^{(i)}}'(u)$, with the source terms $S_{\pm}^{(i)}(u)$ depending on the solutions at lower orders in the perturbative expansion. In principle, exact integral solutions to these equations can be found by using the method of integrating factors. We present the technical details of the perturbative solutions in appendix \ref{app:perturbativedetails}, and focus on the physical consequences in the following.

Using (\ref{eq:rescaledcorrelatorextractions}), the diagonal correlators are 
\begin{equation}
\label{eq:diagcorrholo}
\langle \mathcal{J}_\pm \mathcal{J}_\pm \rangle(\omega)=\frac{\mp\tilde{m}^2r_0^3\gamma_{\pm}}{\left(\gamma_+-\gamma_-\right)}\frac{\tilde{\omega}{F_\pm^{(1)}}'(0)+\tilde{\omega}^2{F_\pm^{(2)}}'(0)+O(\tilde{\omega}^3)}{1+\tilde{\omega}F_\pm^{(1)}(0)+\tilde{\omega}^2F_\pm^{(2)}(0)+O(\tilde{\omega}^3)},
\end{equation}
and these can easily be combined to give the full conductivities using (\ref{eq:condintermsdiagonalcurrents}). The DC conductivities are controlled only by ${F_{\pm}^{(1)}}'(0)$, which can be analytically determined exactly as a function of $T,\mu,m$ (see appendix \ref{app:perturbativedetails}). Using these expressions, we recover the results (\ref{eq:hologDCresults}) for the DC conductivities which were discussed extensively in section \ref{sec:DCconductivitiessection}. This provides an independent check of the horizon formula results of \cite{Andrade:2013gsa,Donos:2014cya} from a Kubo formula calculation.

To understand the physical origin of each contribution to the conductivities, we must determine their frequency dependence by working to higher orders in the perturbative expansion $\tilde{\omega}$. We were not able to obtain analytic results for general $T,\mu,m$ at these higher orders.\footnote{We note that exact results in $m,T$ can be obtained for the neutral $\mu=0$ state.} Instead, we have focused on the limit of slow momentum relaxation $\tilde{m}\ll 1$, and computed the conductivities in a perturbative expansion at small $\tilde{\omega}$ and $\tilde{m}^2$, with $\tilde{\omega}\sim\tilde{m}^2\ll 1$, i.e.~at long timescales, comparable to the momentum lifetime. For some of the coefficients of the terms in (\ref{eq:diagcorrholo}), we have only been able to obtain analytic answers perturbatively in $\tilde{\mu}$. However, our final result for the coherent and incoherent contributions to the DC conductivities will be non-perturbative in $\tilde{\mu}$. The details of the perturbative solutions are given in appendix \ref{app:perturbativedetails}. Note that we are working with dimensionless variables normalised by $r_0$: for $\tilde{m}\ll1$, $r_0\sim T$ when $\mu\ll T$ and $r_0\sim\mu$ when $T\ll\mu$.

\subsection{The incoherent contribution}

In this subsection, we will focus on the $\langle J_+J_+\rangle(\omega)$ correlator at small $\tilde{m}$. Recall that in the strict $m=0$ limit, $J_+$ is the current (\ref{eq:smallmdecoupledcurrents}) which decouples from momentum and which is therefore completely incoherent. Our perturbative calculation, described in appendix \ref{app:perturbativedetails}, yields the $J_+$ conductivity\footnote{\label{footnote:normalisation}We normalise by a factor of $m^{-2}$ due to the ubiquitous appearance of such a factor in (\ref{eq:condintermsdiagonalcurrents}) at small $m$.}
\begin{equation}
\label{eq:JplusJplusresult}
\Sigma^+(\omega)\equiv\frac{1}{m^2}\frac{i}{\omega}\langle \mathcal{J}_+\mathcal{J}_+\rangle(\omega)=\frac{\left[\sigma_Q+O(\tilde{m}^2)\right]+\left[\beta_1+O(\tilde{m}^2)\right]\tilde{\omega}+O(\tilde{\omega}^2)}{1+\left[\beta_2+O(\tilde{m}^2)\right]\tilde{\omega}+\left[\beta_3+O(\tilde{m}^2)\right]\tilde{\omega}^2+O(\tilde{\omega}^3)},
\end{equation}
where
\begin{equation}
\begin{aligned}
&\beta_1=-\frac{i}{18}\left(\sqrt{3}\pi+9\log3\right)-\frac{i\tilde{\mu}^2}{216}\left(5\sqrt{3}\pi-63\log3\right)+\frac{i\tilde{\mu}^4}{648}\left(7\sqrt{3}\pi-72\log3\right)+O(\tilde{\mu}^6),\\
&\beta_2=-\frac{i}{18}\left(\sqrt{3}\pi+9\log3\right)+\frac{i\tilde{\mu}^2}{216}\left(19\sqrt{3}\pi-9\log3\right)-\frac{5i\tilde{\mu}^4}{72}\left(\sqrt{3}\pi-4\log3\right)+O(\tilde{\mu}^6),\\
&\beta_3=-\frac{1}{216}\left(\pi^2+6\sqrt{3}\pi\log3-27\left(\log3\right)^2\right)+O(\tilde{\mu}^2).
\end{aligned}
\end{equation}
The nature of the transport of the current $J_+$ is encoded in its pole structure: coherent transport is caused by a parametrically long-lived excitation. In our system, this would be a Drude-like excitation due to the slow relaxation of momentum, which has the dispersion relation $\tilde{\omega}\sim-i\tilde{m}^2$ \cite{Davison:2013jba,Davison:2014lua}.

It is clear from (\ref{eq:JplusJplusresult}) that there is no such long lived excitation transporting the current $J_+$. Our calculation shows that the longest lived collective excitations in this sector have microscopic lifetimes $\sim\tilde{m}^0$.\footnote{We cannot give quantitative results for the lifetime, as frequencies $\tilde{\omega}\sim 1$ are outside of the range of validity of our perturbative calculation.} Therefore the contributions of $\langle J_+J_+\rangle$ to the thermoelectric conductivities (\ref{eq:condintermsdiagonalcurrents}) are all incoherent. To the order of the perturbative expansion to which we are working, the contribution of these incoherent processes to the full thermoelectric conductivities of the system is 
\begin{equation}
\label{eq:pluscontributionstoeachconductivity}
\begin{aligned}
&\sigma^+(\omega)=\sigma_Q+O(\tilde{\omega},\tilde{m}^2)\,,\qquad \alpha^+(\omega)=-\frac{\mu}{T}\sigma_Q+O(\tilde{\omega},\tilde{m}^2),\\
&\bar{\kappa}^+(\omega)=\frac{\mu^2}{T}\sigma_Q+O(\tilde{\omega},\tilde{m}^2)\,,\qquad \sigma_Q=\frac{\left(12-\tilde{\mu }^2\right)^2}{9 \left(4+\tilde{\mu }^2\right)^2}.
\end{aligned}
\end{equation}
This is entirely in agreement with the incoherent contributions to the conductivities predicted in (\ref{eq:hydroacresult}) by the hydrodynamic effective theory of \cite{Hartnoll:2007ih}.

\subsection{The coherent contribution \label{sec:coherentresultsholo}}

We now turn to the $\langle J_-J_-\rangle(\omega)$ correlator. We know that this must have a coherent component at small $m$ since the conductivities themselves do. However it is not clear, a priori, if $J_-$ is transported completely coherently (to the order in $m^2$ to which we are working), or whether it has both coherent and incoherent components. Our perturbative calculation, described in appendix \ref{app:perturbativedetails}, yields the $J_-$ conductivity\footnote{See footnote \ref{footnote:normalisation}.}
\begin{equation}
\label{eq:exactminuscorrelator}
\Sigma^-(\omega)\equiv\frac{1}{m^2}\frac{i}{\omega}\langle \mathcal{J}_-\mathcal{J}_-\rangle(\omega)=\frac{\left[a_1+a_2\tilde{m}^2+O(\tilde{m}^4)\right]-\left[b_1+O(\tilde{m}^2)\right]i\tilde{\omega}+O(\tilde{\omega}^2)}{\tilde{m}^2-\left[c_1+c_2\tilde{m}^2+O(\tilde{m}^4)\right]i\tilde{\omega}+\left[d_1+O(\tilde{m}^2)\right]\tilde{\omega}^2+O(\tilde{\omega}^3)},
\end{equation}
where
\begin{equation}
\begin{aligned}
\label{eq:condcoefficientsforourtheory}
&a_1=\tilde{\mu}^2,\\
&a_2=\frac{8\tilde{\mu}^2\left(12+\tilde{\mu}^2\right)}{9\left(4+\tilde{\mu}^2\right)^2},\\
&b_1=\frac{\tilde{\mu}^4}{54}\left(\sqrt{3}\pi-18+9\log3\right)+\frac{\tilde{\mu}^6}{216}\left(27-4\sqrt{3}\pi-6\log3\right)+O(\tilde{\mu}^8),\\
&c_1=\frac{3}{4}\left(4+\tilde{\mu}^2\right),\\
&c_2=-\frac{1}{18}\left(9+\sqrt{3}\pi-9\log3\right)+\frac{\tilde{\mu}^2}{216}\left(72-\sqrt{3}\pi+9\log3\right)+\frac{\tilde{\mu}^4}{864}\left(\sqrt{3}\pi-84+3\log3\right)+O(\tilde{\mu}^6),\\
&d_1=\tilde{\mu}^2\left(1-\frac{\pi}{6\sqrt{3}}-\frac{\log3}{2}\right)+\frac{\tilde{\mu}^4}{24}\left(-3+\sqrt{3}\pi-\log3\right)+O(\tilde{\mu}^6).
\end{aligned}
\end{equation}
Note that we are working to the first subleading order in the small $\tilde{\omega}\sim\tilde{m}^2$ expansion in both the numerator and the denominator.

\begin{figure}
\begin{tabular}{c}
\includegraphics[width=.8\textwidth]{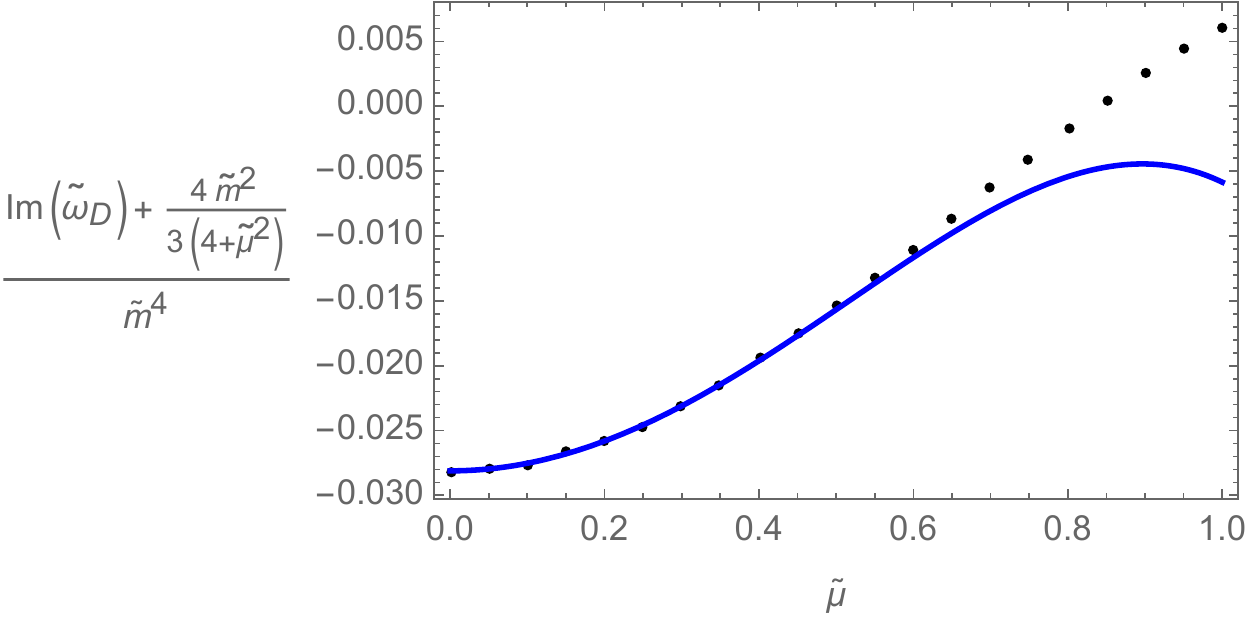}
\end{tabular}
\caption{A comparison of our analytic result (\ref{eq:holodrudepolelocation}) for the $O(\tilde{m}^4)$ correction to the location of the Drude-like pole (blue line), and the exact location obtained numerically for $\tilde{m}=1/10$ (black dots). There is excellent agreement at small $\tilde{\mu}$ where our perturbative analytic result should be accurate.} 
\label{fig:poledispersioncomparison}
\end{figure}

As expected, this correlator has a pole at $\tilde{\omega}\sim-i\tilde{m}^2$. This pole corresponds to the existence of a parametrically long-lived Drude-like excitation due to the slow relaxation of momentum, and will give coherent contributions to the thermoelectric conductivities. Our perturbative calculation allows us to determine subleading corrections to the location of this Drude-like pole $\tilde{\omega}_D$
\begin{equation}
\begin{aligned}
\label{eq:holodrudepolelocation}
\tilde{\omega}_D=-i\Biggl[&\frac{4}{3\left(4+\tilde{\mu}^2\right)}\tilde{m}^2+\Biggl\{\frac{1}{162}\left(9+\sqrt{3}\pi-9\log3\right)+\frac{\tilde{\mu}^2}{1944}\left(-198-\sqrt{3}\pi+81\log3\right)\\
&+\frac{\tilde{\mu}^4}{2592}\left(187-6\sqrt{3}\pi-54\log3\right)+O(\tilde{\mu}^6)\Biggr\}\tilde{m}^4+O(\tilde{m}^6)\Biggr].
\end{aligned}
\end{equation}
The $O(\tilde{m}^2)$ term agrees with that of \cite{Davison:2013jba}, as it should (given the similarities between the gravity theories under study here and there \cite{Andrade:2013gsa}). The $\tilde\mu=0$ limit of this term is also in agreement with \cite{Davison:2014lua}. At $O(\tilde{m}^4)$, we could only calculate the location of the pole perturbatively in $\tilde{\mu}$, as is clear from the result (\ref{eq:holodrudepolelocation}). As a check of our calculation, a comparison of this analytic result with a numerical calculation of the pole location is shown in figure \ref{fig:poledispersioncomparison}: there is excellent agreement for appropriately small values of $\tilde{\mu}$. For the neutral $\mu=0$ case, interestingly, the location of the pole to $O(\tilde{m}^4)$ is equivalent to that of the transverse momentum diffusion pole in the translationally invariant system dual to the Schwarzschild-AdS$_4$ black brane (see table III of \cite{Natsuume:2007ty}), after replacing $m$ with the wavenumber $q$.

From this pole, we can identify the momentum relaxation rate $\Gamma$ as $\Gamma=ir_0\tilde{\omega}_D$, which can be written (perhaps more illuminatingly) as
\begin{equation}
\label{eq:GammaExp2}
\begin{aligned}
\Gamma&=\frac{s m^2}{4\pi\left(\epsilon+p\right)}\left[1+\lambda m^2+O(m^4)\right],\\
\lambda&=\frac{\sqrt{3}\pi-9\log3}{96\pi^2T^2}+\frac{9\mu^2\left(\log3-2\right)}{256\pi^4 T^4}-\frac{9\mu^4\left(42\log3+5\sqrt{3}\pi-132\right)}{32768\pi^6 T^6}+O\left(\frac{\mu^6}{T^8}\right),
\end{aligned}
\end{equation}
where the thermodynamic quantities are $m$-dependent. There will, of course, be other poles in the correlator with decay rates $\sim\tilde{m}^0$, but our perturbative calculation is not able to accurately capture these.

The current $J_-$ clearly has a coherent contribution to its transport, due to the existence of the Drude-like pole. Generically, we would also expect there to be an incoherent component to its transport. To quantify this, we calculate the residue of the Drude-like pole in the conductivity of $J_-$:
\begin{equation}
Z_D\equiv\lim_{\tilde{\omega}\rightarrow\tilde{\omega}_D}\left(\tilde{\omega}-\tilde{\omega}_D\right)\Sigma^-(\tilde{\omega}).
\end{equation}
The natural definition of the contribution of the coherent excitation to $\Sigma^-_{DC}$ is then $-Z_{D}/\tilde{\omega}_D$. For a conductivity of the form (\ref{eq:exactminuscorrelator}), this yields
\begin{equation}
\begin{aligned}
\label{eq:generalcoherentcontribut}
-\frac{Z_D}{\tilde{\omega}_D}&=\frac{a_1}{\tilde m^2}+a_2-\left(\frac{b_1}{c_1}+\frac{a_1 d_1}{c_1^2}\right)+O(\tilde{m}^2)\\
&=\Sigma^-_{DC}-\left(\frac{b_1}{c_1}+\frac{a_1 d_1}{c_1^2}\right)+O(\tilde{m}^2)
\end{aligned}
\end{equation}
as the coherent contribution to $\Sigma^-_{DC}$. At order $\tilde{m}^{-2}$, all of the DC conductivity comes from the Drude peak, but at the first subleading order, $O(\tilde{m}^0)$, this is not necessarily the case. The term in brackets in (\ref{eq:generalcoherentcontribut}) indicates a part of $\Sigma_{DC}^-$ which does \textit{not} come from the Drude peak i.e.~it is an incoherent contribution. However, substituting in the explicit expressions for $a_1,b_1,c_1$ and $d_1$ for our system (\ref{eq:condcoefficientsforourtheory}), this potential incoherent component of the DC conductivity vanishes identically! This indicates that, to the order in $\mu$ to which our result (\ref{eq:condcoefficientsforourtheory}) is valid, the entire DC conductivity of $J_-$ comes from the Drude-like excitation, up to and including the first subleading order in the $\tilde{m}^2$ expansion.

In fact, we can show that this is true to all orders in $\tilde{\mu}$. Although we do not know individually how $a_1,b_1,c_1$ and $d_1$ depend upon $\tilde{\mu}$, it is easy to check that the precise combination appearing in the brackets in equation (\ref{eq:generalcoherentcontribut}) must vanish, by demanding that in the strict $m=0$ limit, we reproduce the hydrodynamic results of \cite{Hartnoll:2007ip}. This assumption of continuity of the hydrodynamic limit is manifestly true up to $O(\tilde{\mu}^4)$, and we believe it should be true to all orders. In summary: up to and including the first subleading order in the $\tilde{m}^2$ expansion, the entire DC conductivity of $J_-$ comes from the Drude-like excitation -- there is no incoherent component at this order. At higher orders in the $\tilde{m}^2$ expansion, we expect $\Sigma^-_{DC}$ to be a sum of both coherent and incoherent contributions.

The contributions of this sector to each conductivity are therefore totally coherent, to this order, and given by
\begin{equation}
\label{eq:minuscontributionstoeachconductivity}
\begin{aligned}
\sigma^-\left(\omega\right)&=\frac{\frac{\mu^2}{m^2}+\left(1-\sigma_Q\right)+O(\tilde{\omega},\tilde{m}^2)}{1-i\omega/\Gamma}+O(\tilde{\omega},\tilde{m}^2),\\
\alpha^-\left(\omega\right)&=\frac{\frac{4\pi n}{m^2}+\frac{\mu}{T}\sigma_Q+O(\tilde{\omega},\tilde{m}^2)}{1-i\omega/\Gamma}+O(\tilde{\omega},\tilde{m}^2),\\
\kappa^-\left(\omega\right)&=\frac{\frac{4\pi sT}{m^2}-\frac{\mu^2}{T}\sigma_Q+O(\tilde{\omega},\tilde{m}^2)}{1-i\omega/\Gamma}+O(\tilde{\omega},\tilde{m}^2),
\end{aligned}
\end{equation}
where $\sigma_Q$ is given in (\ref{eq:rnads4sigmaq}) and $\Gamma$ is given in (\ref{eq:GammaExp2}). At higher orders in the small $\tilde{\omega},\tilde{m}^2$ expansion, we expect that the contributions of $J_-$ to the conductivities will be comprised of both coherent and incoherent pieces.

\subsection{Discussion}

Collecting the results (\ref{eq:pluscontributionstoeachconductivity}) and (\ref{eq:minuscontributionstoeachconductivity}), the thermoelectric conductivities are given by equations (\ref{eq:OurResults}) in the limit of slow momentum relaxation. As we have demonstrated, the coherent part of each conductivity comes solely from $J_-$ at this order, while the incoherent part comes only from $J_+$. For an easier comparison with the memory matrix and hydrodynamic formulae (\ref{eq:memmatrixacresults}) and (\ref{eq:hydroacresult}), we can change variables from $m^2$ to $\Gamma$ and write the conductivities to subleading order in a small $\omega\sim\Gamma$ expansion
\begin{equation}
\label{eq:section3results}
\begin{aligned}
\sigma\left(\omega\right)&=\frac{\frac{n^2}{\epsilon+p}+\Gamma\left(1-\sigma_Q+\lambda\mu^2\right)+O(\Gamma^2,
\omega\Gamma,\omega^2)}{\Gamma-i\omega}+\sigma_Q+O(\omega,\Gamma),\\
\alpha\left(\omega\right)&=\frac{\frac{n s}{\epsilon+p}+\Gamma\left(\frac{\mu}{T}\sigma_Q+4\pi n\lambda\right)+O(\Gamma^2,
\omega\Gamma,\omega^2)}{\Gamma-i\omega}-\frac{\mu}{T}\sigma_Q+O(\omega,\Gamma),\\
\bar{\kappa}\left(\omega\right)&=\frac{\frac{s^2T}{\epsilon+p}+\Gamma\left(-\frac{\mu^2}{T}\sigma_Q+4\pi s T\lambda\right)+O(\Gamma^2,
\omega\Gamma,\omega^2)}{\Gamma-i\omega}+\frac{\mu^2}{T}\sigma_Q+O(\omega,\Gamma).
\end{aligned}
\end{equation}
In the translationally invariant limit $\Gamma=0$, these agree with the results of \cite{Hartnoll:2007ip}, and the hydrodynamic formulae (\ref{eq:hydroacresult}). When $\Gamma\ne0$, they agree with the memory matrix results (\ref{eq:memmatrixacresults}) and hydrodynamic results (\ref{eq:hydroacresult}) at leading order in the small $\omega,\Gamma$ expansion, but not at subleading order. The subleading corrections in (\ref{eq:section3results}) are comprised of two independent pieces: an incoherent contribution, and a coherent contribution (a correction to the weight of the Drude peak). The hydrodynamic results (\ref{eq:hydroacresult}) correctly capture the incoherent contribution but not the correction to the Drude peak. The memory matrix results (\ref{eq:memmatrixacresults}) can also be extended to include an incoherent contribution \cite{Lucas:2015pxa}, but not yet the correction to the Drude peak. Since the subleading correction to the Drude peak enters at the same order (in a small $\Gamma$ or small $m$ expansion) as the incoherent contribution in our holographic theory, it is important that these effective theories are extended to incorporate this correction to the Drude peak.

In the limit of zero chemical potential, the conductivities are given by
\begin{equation}
\begin{aligned}
\label{eq:OurResults2}
\sigma\left(\omega\right)&=1,\\
\alpha\left(\omega\right)&=0,\\
\bar{\kappa}\left(\omega\right)&=\frac{\frac{s^2T}{\epsilon+p}+4\pi s T\lambda\Gamma}{\Gamma-i\omega},
\end{aligned}
\end{equation}
to this order. The electric conductivity $\sigma$ is totally incoherent since $J$ decouples from $P$ when $\mu=0$, while the thermal conductivity $\bar{\kappa}$ is totally coherent at this order, confirming further the results of \cite{Davison:2014lua}.

In the introduction, we noted that previous works have tried to identify the $m$-independent contribution to $\sigma_{DC}$ (which numerically is equal to 1) as being the incoherent component of the electrical conductivity. As is clear from our results, this is not the case. However, the $m$-independent contribution in this theory \textit{can} be identified as being the DC value of the electrical current in the absence of heat flow \cite{Donos:2014cya}
\begin{equation}
\label{eq:sigma0Q}
\left.\sigma_{DC}\right|_{Q=0}=\sigma_{DC}-\frac{T\alpha_{DC}^2}{\bar{\kappa}_{DC}}=1.
\end{equation}
This result can be generalised to more complicated holographic theories in a natural way \cite{Donos:2014cya,Donos:2014yya}. With our results, we can revisit this computation and determine how this conductivity depends on frequency, finding 
\begin{equation}
\begin{aligned}
\sigma(\omega)\bigr|_{Q=0}&=\frac{\Gamma}{\Gamma-i\omega}\left[1-\frac{\left(\epsilon+p\right)^2}{s^2T^2}\sigma_Q+O\left(\omega,\Gamma\right)\right]+\frac{\left(\epsilon+p\right)^2}{s^2T^2}\sigma_Q+O\left(\omega,\Gamma\right),\\
&=1+O\left(\Gamma,\omega,\frac{\Gamma^2}{\Gamma-i\omega},\ldots\right),
\end{aligned}
\end{equation}
to the order to which our calculations are valid and recalling the value of $\sigma_Q$ \eqref{eq:rnads4sigmaq}. This conductivity is totally incoherent to this order, and does not have any contributions from subleading corrections to the Drude peak. It would be interesting to determine whether this is also the case to higher order in the expansion, or whether $\sigma(\omega)\bigr|_{Q=0}$ has contributions $\sim\Gamma^2/(\Gamma-i\omega)$ etc. Similarly, the heat conductivity in the absence of electrical current is 
\begin{equation}
\begin{aligned}
\bar{\kappa}(\omega)\bigr|_{J=0}&=\frac{\Gamma}{\Gamma-i\omega}\left[\frac{Ts^2}{n^2}-\frac{\left(\epsilon+p\right)^2}{n^2T}\sigma_Q+O\left(\omega,\Gamma\right)\right]+\frac{\left(\epsilon+p\right)^2}{n^2T}\sigma_Q+O\left(\omega,\Gamma\right),\\
&=\frac{s^2T}{n^2}+O\left(\Gamma,\omega,\frac{\Gamma^2}{\Gamma-i\omega},\ldots\right),
\end{aligned}
\end{equation}
which is totally incoherent to the order to which we are working. We note that the absence of any leading order contribution $\sim\Gamma^0/\left(\Gamma-i\omega\right)$ to these conductivities is as expected from \cite{Mahajan:2013cja}.

\section{Outlook \label{section:outlook}}

We have shown that the transport of heat and charge in the state with momentum relaxation, dual to (\ref{eq:axiontheorymetricsolution}), can naturally be expressed in terms of the two currents $J_\pm$, given in \eqref{eq:defofdecoupledcurrents}, which diagonalise the thermoelectric conductivity matrix \eqref{eq:transportmatrixintroduction} for all values of $m$. In certain limits, the form of these currents can be used to understand the physical processes underpinning the transport properties. In the limit of very fast momentum relaxation ($m\rightarrow\infty$), the heat and electrical currents decouple, as they do in the charge conjugation symmetric limit. In the limit of no momentum relaxation ($m=0$), the decoupled currents are the coherent energy current $J_E$ (which is equal to the total momentum $P$), and the current $J_E-\frac{3\epsilon_0}{2n_0}J$ which is completely incoherent, as it decouples from the total momentum $P$ \cite{DGH}.

We have analytically computed the low frequency behaviour of the conductivities in the limit of slow momentum relaxation (small $m$). In this limit, the decoupled currents $J_\pm$ are still controlled by qualitatively different physical processes. To the first subleading order at small $m$, $J_-$ remains coherent, i.e.~it is controlled by the momentum relaxation timescale of the system, while $J_+$ remains incoherent, i.e.~it is controlled by the intrinsic relaxation timescale of the system. There is a smooth $m\rightarrow0$ limit. The two independent contributions combine in a very non-trivial way to form the DC conductivities \eqref{eq:hologDCresults} --- it is not easy to guess how the DC formul\ae\ should be divided up into coherent and incoherent contributions without any other information.

Our results highlight the fact that subleading corrections to the Drude weight enter at the same order (in $m$) as the leading incoherent contribution to each thermoelectric conductivity. The apparent discrepancies between the holographic DC conductivities and those of the memory matrix or hydrodynamic descriptions are due to the neglection of corrections to the Drude weight in these effective theories.

There are several directions which are worth pursuing further:

\paragraph{Spatially resolved transport}

We have considered the transport of the spatially uniform components of the charges and currents. A natural extension would be to study the transport of the non-zero wavenumber $k$ harmonics, to understand how charge is transported over different distance scales. In the limit of slow momentum relaxation, we expect that, at low energies, $J_-$ will be transported by sound at short distances (large $k$) and diffusion at long distances (small $k$), as was observed in \cite{Davison:2014lua} for a zero density system. In contrast to this, we expect that $J_+$ will be transported by diffusion at all distance scales, due to its incoherent nature.

\paragraph{Magnetotransport}

Building on \cite{Hartnoll:2007ih}, a number of recent articles have revisited the problem of magnetotransport with momentum relaxation by computing the thermoelectric and Hall conductivities either holographically \cite{Blake:2014yla,Amoretti:2015gna,Blake:2015ina,Kim:2015wba} or with memory matrices \cite{Lucas:2015pxa}. To resolve the discrepancy between the hydrodynamic, memory matrix, and holographic DC calculations, it would be worthwhile to adapt our techniques to calculate the frequency dependent conductivities at non-zero $B$. Extending our calculations to non-zero $B$ would also allow us to examine whether the Hall angle receives contributions from both coherent and incoherent processes and how this relates to the interpretation of its temperature scaling in terms of two timescales \cite{Blake:2014yla}.

It was recently proposed \cite{Hayes:2014} that the timescale setting the resistivity scaling of the strange metallic region of a certain iron pnictide compound is proportional to the square root of a sum of squares of the temperature and magnetic field. With this in mind, it would be very interesting to determine the dependence of the appropriate timescale (momentum relaxation rate or diffusion constant) on the magnetic field in holographic systems. This could be done by adapting the methods we have used here.

\paragraph{More general theories}

It would be very worthwhile to extend our work to more general holographic theories with slow momentum relaxation, in which a hydrodynamic limit exists at non-zero temperatures. This should be the case when a Drude-like pole dominates the correlators at sufficiently low energy scales. Holographic theories can exhibit branch cut formation in the $T\rightarrow0$ limit, due to a coalescense of poles with decay rates differing by $\sim T$.  Although our analysis will not capture these poles, a hydrodynamic limit should be valid when $\omega,\Gamma\ll T$, as this is when the Drude-like excitation is parametrically longer lived than the rest.

In theories where there is a neutral scalar which can run logarithmically in the interior of the geometry \cite{Donos:2014uba,Gouteraux:2014hca,Donos:2014oha}, we would expect that the temperature scalings of the coherent and incoherent contributions to the conductivities can be different from one another. With this additional hierarchy of scales, it may be possible to find states with slow momentum relaxation where the effects of the incoherent contribution are parametrically larger (or smaller) than corrections to the Drude peak. The method used in \cite{Lucas:2015vna} may be useful for more general theories.

Another question is the sensitivity of our results to the choice of momentum relaxation mechanism: would they be modified if we had instead used random-field disorder \cite{Davison:2013txa,Lucas:2014zea,Lucas:2015vna,Hartnoll:2014cua,O'Keeffe:2015awa,Hartnoll:2015faa}, or homogeneous \cite{Donos:2012js,Donos:2014oha,Donos:2014gya} or inhomogeneous lattices \cite{Horowitz:2012ky,Blake:2013owa,Donos:2014yya} to break translational invariance? Furthermore, if we had broken translational invariance with electrically charged, rather than neutral, operators, would this affect the nature of transport in the system? In particular, we interpreted the decoupling of $J$ and $Q$ at large $m$ as being a consequence of the state's thermodynamics becoming dominated by the neutral scalar degrees of freedom. Does the same decoupling occur (at low frequencies) when these neutral operators are not present?

A qualitatively different class of holographic systems with finite conductivities are probe brane systems, whose DC electrical conductivity can  be written as the square root of the sum of two terms \cite{Karch:2007pd}, one of which is often interpreted as a `Drude-like' term (and can also be computed from the drag force on the charge carriers), the other as a `pair creation' term. It would be interesting to verify this interpretation by analytically computing the low frequency, linear response conductivity in such a system, as we have done here.

\paragraph{Effective theories of thermoelectric transport with slow momentum relaxation}

Our computation has highlighted what needs to be done to refine existing effective hydrodynamic \cite{Hartnoll:2007ih} or memory matrix \cite{Lucas:2015pxa} theories of transport in the presence of slow momentum relaxation, such that they are consistent with the holographic computations of DC conductivities. These effective theories should be extended to take into account order $\Gamma$ corrections to the weight of the Drude peak. These produce $O(\Gamma^0)$ corrections to the DC conductivities, which are the same order as the incoherent $\sigma_Q$ contributions. This is an excellent example of how gauge/gravity duality can contribute to the understanding of transport in strongly correlated systems in general, by providing a consistent and reliable framework from which effective theories can be extracted, or to which effective theories can be compared.

\acknowledgments
We are grateful to Mike Blake, Sa\v{s}o Grozdanov, Sean Hartnoll and Jan Zaanen for many helpful and insightful discussions. We thank Mike Blake, Sean Hartnoll, Elias Kiritsis, Andy Lucas and Subir Sachdev for comments on a draft of this manuscript. We also thank the Galileo Galilei Institute for Theoretical Physics (GGI) for hospitality, and the INFN for partial support, during the completion of this work. These results were presented at the Gauge/Gravity Duality 2015 conference at the GGI (Florence, Italy) in April 2015. The work of R.D. is supported by a VIDI grant from NWO, the Netherlands Organisation for Scientific Research. The work of B.G. is supported by the Marie Curie International Outgoing Fellowship nr 624054 within the 7th European Community Framework Programme FP7/2007-2013. 

\appendix

\section{The on-shell action \label{app:onshellactionappendix}}

The on-shell action of the theory (\ref{eq:axiontheoryaction}), to quadratic order in the perturbations (\ref{eq:quadraticperturbationsdefinition}) around the solution (\ref{eq:axiontheorymetricsolution}), is
\begin{equation}
\begin{aligned}
S=\int d^2x\frac{d\omega}{2\pi}\Biggl\{&\frac{3m}{2\left(\omega^2-m^2\right)}{h^y_t}^{(0)}(-\omega)\left[m{h^y_t}^{(3)}(\omega)+i\omega{\chi_2}^{(3)}(\omega)\right]+\frac{1}{2}{a_y}^{(0)}(-\omega){a_y}^{(1)}(\omega)\\
&-\frac{r_0\left(\mu^2+4r_0^2-2m^2\right)}{4}{h^y_t}^{(0)}(-\omega){h^y_t}^{(0)}(\omega)-\frac{r_0\mu\left(2\omega^2-m^2\right)}{2\left(\omega^2-m^2\right)}{h^y_t}^{(0)}(-\omega)a_y^{(0)}(\omega)\Biggr\},
\end{aligned}
\end{equation}
where we have expanded a generic field perturbation $\delta\varphi(r,\omega)$ near the boundary as
\begin{equation}
\delta\varphi(r,\omega)=\sum_n \frac{{\delta\varphi}^{(n)}(\omega)}{r^{n}},
\end{equation}
and set the scalar operator source term $\chi_2^{(0)}(\omega)$ to zero. From this, we can use the standard AdS/CFT dictionary \cite{Son:2002sd} to calculate expressions for the retarded Green's functions of the operators dual to each field perturbation, in terms of $a_y^{(0)}(\omega),a_y^{(1)}(\omega)$, etc. These can then be rewritten in terms of the near-boundary expansions of the decoupled variables using their definitions (\ref{eq:defnofdecoupledbulkfields}). To compute the subtracted correlators (\ref{eq:subtractedcorrelatordefinition}) that enter in the Kubo formulae (\ref{eq:thermoelecgreensfns}) for the conductivities, we must subtract the retarded Green's functions when $\omega=0$ and $k\rightarrow0$, where $k$ is the wavenumber of the perturbation in the $y$-direction. These were obtained by computing the on-shell action for fluctuations of this kind, yielding the expressions (\ref{eq:subtractedbasiccorrs}) for the subtracted correlators. A non-trivial consistency check of our calculations (including contact terms) is that, after solving the equations of motion and substituting these solutions into (\ref{eq:subtractedbasiccorrs}), we find that the conductivities are free of $i/\omega$ poles, as should be the case on physical grounds.

\section{Details of the perturbative calculations \label{app:perturbativedetails}}

In this appendix, we give details of the perturbative solutions for the functions $\mathcal{F}_{\pm}(u)$, defined in (\ref{eq:decoupledansatzeholo}). At leading order in $\tilde{\omega}$, the solutions which obey the correct boundary conditions at the horizon are simply constants $F_\pm^{(0)}(u)=C_{\pm}$. The value of these constants is unimportant and will cancel out in the final answers for the conductivities, and so for convenience we set $C_\pm=1$.

At $O(\tilde{\omega})$ in the expansion, we can formally write the solutions as integrals
\begin{equation}
\label{eq:firstorderintegral}
F_\pm^{(1)}(u)=\int^u_1dx\frac{-4\left(\tilde{m}\gamma_\pm+\tilde{\mu}\right)^2\left(\tilde{\mu}^2+2\tilde{m}^2-12\right)+4xh_\pm(x)\left[\tilde{m}^2\left(6x-4\right)+x\left(\tilde{\mu}^2\left(4x-3\right)-12\right)\right]}{i\left(x-1\right)h_\pm(x)\left(\tilde{\mu}^2+2\tilde{m}^2-12\right)\left[-4-4x+x^2\left(-4+2\tilde{m}^2+x\tilde{\mu}^2\right)\right]},
\end{equation}
where
\begin{equation}
h_\pm(x)=\left(\tilde{m}\gamma_\pm+x\tilde{\mu}\right)^2.
\end{equation}
From these integrals, it is straightforward to analytically calculate the constants ${F_{\pm}^{(1)}}'(0)$ that control the DC conductivities
\begin{equation}
{F_{\pm}^{(1)}}'(0)=i\frac{\left(\tilde{m}\gamma_{\pm}+\tilde{\mu}\right)^2}{\tilde{m}^2\gamma_{\pm}^2}.
\end{equation}
However, we could not do the integrals analytically and find exact expressions for ${F_{\pm}^{(1)}}(0)$. We are primarily interested in the small $\tilde{m}$ limit of the conductivities, and thus the small $\tilde{m}$ limit of the integrals. For $F_+^{(1)}(0)$, it is straightforward to expand the integrand at small $\tilde{m}$, and find that the leading order term is of order $\tilde{m}^0$. This means that there is no Drude-like excitation in the conductivity of $J_+$, which is therefore incoherent. For our purposes, this is all we need to know. For completeness, we note that it is not possible to integrate the leading term of the integrand analytically for general $\tilde{\mu}$, but that it is possible in a small $\tilde{\mu}$ expansion:
\begin{equation}
\begin{aligned}
F_+^{(1)}(0)=\Biggl[&-\frac{i}{18}\left(\sqrt{3}\pi+9\log3\right)+\frac{i\tilde{\mu}^2}{216}\left(19\sqrt{3}\pi-9\log3\right)+O(\tilde{\mu}^4)\Biggr]+O(\tilde{m}^2).
\end{aligned}
\end{equation}

For $F_-^{(1)}(0)$, it is more complicated. The small $\tilde{m}$ limit of the integrand is singular due to the form of the function $h_-(x)\sim(\tilde{m}^2+x)^2$ in the denominator: the limits $\tilde{m}\rightarrow0$ and $x\rightarrow0$ do not commute. If we first send $\tilde{m}^2\rightarrow0$, the integrand diverges when $x\rightarrow0$. To correctly evaluate the small $\tilde{m}$ limit, we must take it small but non-zero, so that we accurately include the contribution from integrating over the region $0<x<\tilde{m}^2$. To do this, we change the integration variable to $x=\tilde{m}^2y$ before expanding the integrand at small $\tilde{m}$ and integrating the leading term in this expansion to give
\begin{equation}
\label{eq:minusfirstordersmallmresult}
F_-^{(1)}(0)=-\frac{3i\left(4+\tilde{\mu}^2\right)}{4\tilde{m}^2}+O(\tilde{m}^0).
\end{equation} 
This change of variables is only useful for giving us the leading term: it does not allow us to accurately extract any of the subleading terms in $\tilde{m}$. We have checked that this technique is reliable by explicitly doing the integral numerically and comparing it to our result (\ref{eq:minusfirstordersmallmresult}). The consistency between our analytic pole location and the exact one determined numerically (see figure \ref{fig:poledispersioncomparison}) is also a check of this. The first correction to (\ref{eq:minusfirstordersmallmresult}) is given below in (\ref{eq:firstcorrectiontof1minus}).

At second order in the small $\tilde{\omega}$ expansion, things are even more complicated and we can only get analytic results for $F^{(2)}_\pm(u)$ by performing a double expansion at small $\tilde{m}$ and small $\tilde{\mu}$. The strategy is as follows: we expand the integrand of (\ref{eq:firstorderintegral}) to the second subleading order in $\tilde{\mu}$, and integrate each coefficient to obtain an expression for $F_{\pm}^{(1)}(u)$ which is perturbative in $\tilde{\mu}$ and exact in $\tilde{m}$. This enters as a source in the equations of motion for ${F_{\pm}^{(2)}}'(u)$, for which we can write down formal integral solutions which are much too lengthy to include here. We then expand these integrands to the same order in $\tilde{\mu}$ and again integrate term-by-term to obtain expressions for ${F_{\pm}^{(2)}}'(u)$ which are exact in $\tilde{m}$ but perturbative in $\tilde{\mu}$. The final step is to integrate these expressions, but we could not do this analytically, even in the small $\tilde{\mu}$ expansion. Since we are only interested in the leading order behaviour at small $\tilde{m}$, we expanded each term in the small $\tilde{\mu}$ expansion of the integrand to the lowest order in $\tilde{m}$. For $F_-^{(2)}(u)$, this again was preceded by a rescaling of the integration variable $x=\tilde{m}^2y$ due to the singularity of the $\tilde{m}\rightarrow0$ limit of the integrand. The results are as follows
\begin{align}
\nonumber
&{F^{(2)}_+}'(0)=\left[\frac{1}{18}\left(\sqrt{3}\pi+9\log3\right)+\frac{\tilde{\mu}^2}{216}\left(5\sqrt{3}\pi-63\log3\right)+O(\tilde{\mu}^4)\right]+O(\tilde{m}^2),\\
\nonumber&{F^{(2)}_-}'(0)=\frac{1}{\tilde{m}^4}\left[\frac{\tilde{\mu}^2}{6}\left(\sqrt{3}\pi-18+9\log3\right)-\frac{\tilde{\mu}^4}{24}\left(9+2\sqrt{3}\pi-12\log3\right)+O(\tilde{\mu}^6)\right]+O(\tilde{m}^{-2}),\\
\nonumber&{F^{(2)}_+}(0)=\left[-\frac{1}{216}\left(\pi^2+6\sqrt{3}\pi\log3+27\left(\log3\right)^2\right)\right.\\
\nonumber&
\left.+\frac{\tilde{\mu}^2}{432} \left(\pi ^2-9\left(\log3\right)^2+18 \sqrt{3} \pi  \log3+16 \psi ^{(1)}\left(\frac{2}{3}\right)-16\psi ^{(1)}\left(\frac{1}{3}\right)\right)+O(\tilde{\mu}^4)\right]+O(\tilde{m}^2),\\
\nonumber&{F^{(2)}_-}(0)=\frac{1}{\tilde{m}^2}\left[\frac{\tilde{\mu}^2}{18}\left(18-\sqrt{3}\pi-9\log3\right)-\frac{\tilde{\mu}^4}{48}\left(6-2\sqrt{3}\pi+2\log3\right)+O(\tilde{\mu}^6)\right]+O(\tilde{m}^0).
\end{align}
where $\psi^{(n)}(z)$ is the polygamma function.
A byproduct of this analysis is that we obtain the $O(\tilde{m}^0)$ correction to (\ref{eq:minusfirstordersmallmresult}), perturbatively in $\tilde{\mu}$:
\begin{equation}
\label{eq:firstcorrectiontof1minus}
\begin{aligned}
F_-^{(1)}(0)=-\frac{3i\left(4+\tilde{\mu}^2\right)}{4\tilde{m}^2}+\Biggl[&\frac{i}{18}\left(9+\sqrt{3}\pi-9\log3\right)+\frac{i}{216}\left(\sqrt{3}\pi-72-9\log3\right)\tilde{\mu}^2\\
&-\frac{i}{864}\left(-84+\sqrt{3}\pi+3\log3\right)\tilde{\mu}^4+O(\tilde{\mu}^6)\Biggr]+O(\tilde{m}^2).
\end{aligned}
\end{equation}
The calculation we have described is quite complex and involves taking two limits (small $\tilde{\mu}$ and small $\tilde{m}$) which, in principle, may not commute. But there are a number of consistency checks we have performed to make sure the expressions above are correct. For example, the location of the Drude-like pole (\ref{eq:holodrudepolelocation}) is sensitive to the value of ${F^{(2)}_-}(0)$, the final quantity derived in the procedure above, and our analytic expression agrees with the exact numerical result (see figure \ref{fig:poledispersioncomparison}). The form of the conductivities in the $m=0$ limit also depend non-trivially on these coefficients (as described in section \ref{sec:coherentresultsholo}), and we have checked that we recover the correct results in this limit. Finally, where possible we have numerically computed the integrals and checked that the results are consistent with our analytic expressions.

\bibliographystyle{JHEP}
\bibliography{DHCDraft}

\end{document}